# Ultralow loss hollow-core negative curvature fibers with nested elliptical antiresonance tubes

JIALI ZHANG,[1] JIE CAO,[1,*] BOYI YANG,[2] XUESHENG LIU,[2] YANG CHENG,[1] LIQUAN DONG,[1] AND QUN HAO[1,*]

[1]*School of Optics and photonics*, *Beijing Institute of Technology, Beijing*, *100081*, *China*
[2]*Beijing Engineering Research Center of Laser Technology, Beijing University of Technology, Beijing, 100124, China*
*\*Corresponding author: ajieanyn@163.com and qhao@bit.edu.cn*

**Abstract:** Hollow-core negative curvature fibers can confine light within air core and have small nonlinearity and dispersion and high damage threshold, thereby attracting a great deal of interest in the field of hollow core fibers. However, reducing the loss of hollow-core negative curvature fibers is a serious problem. On this basis, three new types of fibers with different nested tube structures are proposed in the near-infrared spectral regions and compared in detail with a previously proposed hollow-core negative curvature fiber. We used finite-element method for numerical simulation studies of their transmission loss, bending loss, and single-mode performance, and then the transmission performance of various structural fibers is compared. We found that the nested elliptical antiresonant fiber 1 has better transmission performance than that of the three other types of fibers in the spectral range of 0.72–1.6 μm. Results show that the transmission loss of the $LP_{01}$ mode is as low as $6.45 \times 10^{-6}$ dB/km at $\lambda = 1.06$ μm. To the best of our knowledge, the record low level of transmission loss of hollow-core antiresonant fibers with nested tube structures was created. In addition, the nested elliptical antiresonant fiber 1 has better bending resistance, and its bending loss was below $2.99 \times 10^{-2}$ dB/km at 5 cm bending radius.

## 1. Introduction

Hollow-core fibers (HCFs) [1] are special fibers whose unique ability to confine light in air has attracted a great deal of interest among the researchers over the past decades. These fibers are used in many areas for high power delivery [2], high-speed data communication [3], ultra-short pulse delivery [4], pulse compression [5], terahertz applications [6], and supercontinuum [7]. The first type of HCFs includes hollow-core photonic bandgap fibers (HC-PBGFs), which are divided into OmniGuide HC-PBGFs [8] and 2D HC-PBGFs [9]. The laser travels close to the speed of light in the quasi-vacuum optical environment; thus, the nonlinear effect and dispersion of light are extremely low [3]. However, the surface scattering loss has become the bottleneck limiting the loss reduction of the HC-PBGFs. Only the improvement of the manufacturing process can reduce the loss caused by the rough surface to break through the loss bottleneck. However, the fabrication process of HC-PBGFs can still be remarkably improved. Therefore, continuous research is insignificant. As a result, the researchers preferred hollow-core negative curvature fibers (HC-NCFs) [10].

The HC-NCFs are obtained by simplifying the cladding structure of the HC-PBGFs into a cladding structure. Its guiding mechanism is the antiresonant reflection optical waveguide [11, 12], which is also called hollow-core antiresonant fibers. Various properties of HC-NCFs are mainly realized by designing the cladding component. The different cladding components are reported in the current literature including single ring, double ring, nested ring, elliptical tubes, conjoined tubes, and nested elliptical [3, 4, 13-20]. As an example, in 2012, Yu F of the University of Bath proposed an HC-NCF similar to an ice cream structure, and the minimum loss of the fiber was 34 dB/km at 3.05 μm [21]. In 2014, Poletti from the University of Southampton proposed and simulated an HC-NCF with the cladding geometry of six node-free nested tubes, which has better performance than the photonic bandgap fiber [13]. In 2015, Habib M S et al. from the Technical University of Denmark simulated and designed an HC-NCF with three adjacent nested antiresonant tubes, and the low simulated loss at 0.0015 dB/km is predicted at 1.06 μm [22]. In 2016, Chaudhuri S et al. from Pennsylvania State University designed a regular and nested HC-NCF with elliptical capillary tubes. Simulation results show that the HC-NCF has lower loss and much broader transmission bandwidth than the photonic bandgap fiber [18]. In 2017, Hasan M I proposed an HC-NCF that has an elliptical nested element in the antiresonant tubes. Numerical simulation shows that despite using only a single elliptical nested tube, ultralow loss is achievable over $\lambda = 0.9$ μm to 1.8 μm [23]. In 2018, Wang Y from Beijing University of Technology proposed a new structural design of HC-NCF, and its lowest loss of 2.0 dB/km at 1512 nm [3] was reduced to below 10 dB/km. In 2020, Jasion G T from the University of Southampton reduced the loss of HC-NCFs to 0.28 dB/km by adding a circular nested tube; the loss was reduced to below 1 dB/km [24]. In 2021, Yang et al. proposed a connecting circle HC-NCF with a low confinement loss of ∼0.004 dB/km at $\lambda = 1.06$ μm [25]. Recently, Shaha et al. reported an anisotropic nested antiresonant fiber that provided a confinement loss of 0.0007 dB/km at $\lambda = 1.06$ μm, which is currently the lowest loss achieved by simulation methods among HC-NCFs [26]. Although the loss of HC-NCFs is gradually decreasing, a large gap from its commercialization still exists. Hence, further research is highly expected to solve this problem.

This article presents a nested elliptical antiresonant fiber 1 (NEARF1) with two elliptical tubes nested within the circular tubes (small elliptical tube away from fiber core) arranged in such a way that form nodeless hollow-core fiber structure. To the best of our knowledge, the proposed NEARF1 can, therefore, offer record low $LP_{01}$ loss in the near-IR spectral regions. In addition, considering the low bending loss, effective dual-mode performance, and relatively simple fiber structure, we believe that the proposed NEARF1 is most likely to be fabricated and applied for transmission applications in the near-IR spectral regions.

## 2. Fiber design

We designed three types of new structures and combined one structure reported in the previous literature to study the influence of HC-NCFs with different structures on the beam transmission loss. We also systematically studied the loss performance of HC-NCFs with different ellipse nested tube structures. The cross-sectional views of the four types of HC-NCFs are shown in Fig. 1, where the black area is silica material and the white area is full of air. The nodes between the cladding tubes cause additional resonance of the transmission bands [27]. Thus, we adopt node-free cladding structures. Our main motivation is to find an HC-NCF design capable of broadband low transmission loss near-IR transmission. The outer circular cladding tube negative curvature inhibits the light leakage through the gap of adjacent tubes. The inner elliptical tubes nested within the circular tubes increase the negative curvature and is responsible for the reduced loss.

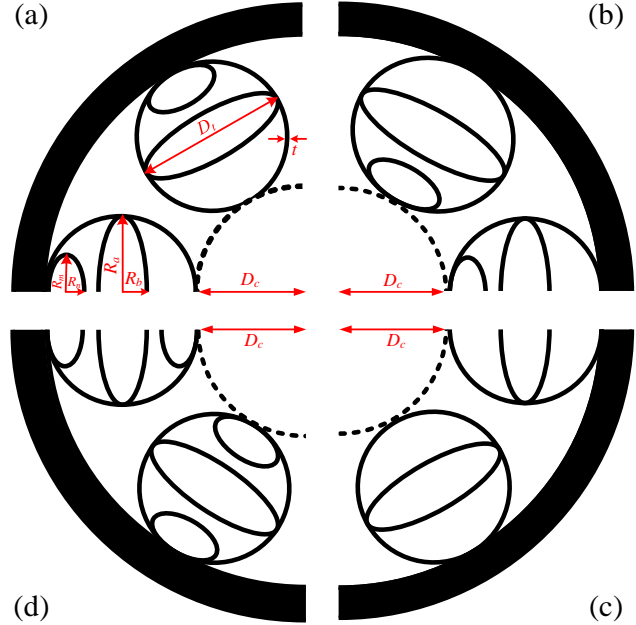

Fig. 1. Geometry includes four quarter structures, namely, (a) nested elliptical antiresonant fiber 1 (NEARF1) with two elliptical tubes (small elliptical tube away from fiber core), (b) nested elliptical antiresonant fiber 2 (NEARF2) with two elliptical tubes (small elliptical tube is near fiber core), (c) nested elliptical antiresonant fiber 3 (NEARF3) with one elliptical tube, and (d) nested elliptical antiresonant fiber 4 (NEARF4) with three elliptical tubes. HC-NCFs have six cladding elements, where the outer circular tube diameter of $D_t$ and the semi-major and semi-minor axis of larger and smaller tubes are $R_a$, $R_b$, $R_m$, and $R_n$. All fibers have the same core radius $D_c$ = 25 μm and uniform silica strut thickness $t$ = 0.35 μm.

The four types of HC-NCFs contain six node-free circular cladding tubes, and the cladding tube has 2, 2, 1, and 3 elliptical nested tubes. The core diameter and cladding tube diameter of the four types of fibers are the same, that is, $2D_c$=50 μm and $D_t$=40 μm, respectively. The so-called core diameter is the maximum diameter of the circle that can be inscribed in the fiber core, as shown in the dotted circle in Fig. 1. The silica strut thickness of the cladding tube and the nested tube are the same, $t$=0.35 μm. The semi-major axis and semi-minor axis of the large ellipse nested tube are $R_a$=19.65 μm and $R_b$=5 μm, respectively. The semi-major axis and semi-minor axis of the small ellipse nested tube are $R_m$= 7 μm and $R_n$=4 μm, respectively. In the full text, the number and position of small elliptical nested tubes are used as variables, and four types of optical fiber structures are simulated to explore their transmission loss, bending loss, and single-mode performance.

## 3. Results and discussion

The full vector finite element method is used to calculate the fiber loss. In Fig. 1, the effective refractive index of air is $n_{air}$=1, and the effective refractive index of silica material is determined by the Sellmeier equation [28].

### 3.1 Loss

The simulation results of the four fiber structures in Fig. 1 are shown in Fig. 2(a). The loss of cladding tubes with different structures are distinguished by different curve colors, and the corresponding mode profile of the core fundamental mode (FM) is shown in Fig. 2(b). The color of the border line of the mode profile corresponds to the color of the loss curve.

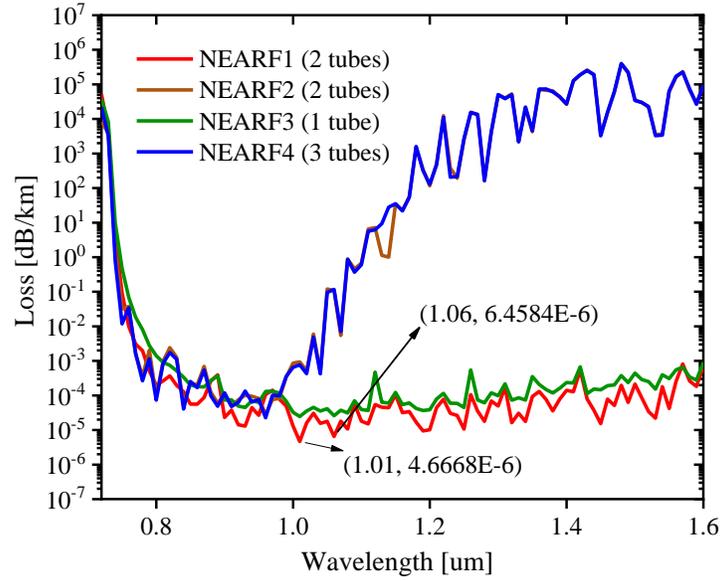

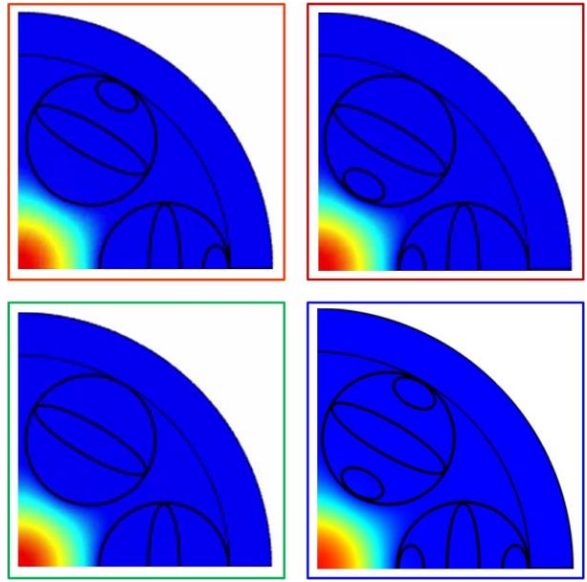

Fig. 2. The loss spectra as function of wavelength for different fiber structures. (a) Loss of NEARF and (b) mode profiles with four cladding structures. The line color of curve relates with the color of frame of geometry.

The loss spectrum in Fig. 2(a) shows that the loss of NEARF1 is lower than that of the three other fibers in the spectral range of 0.72–1.6 μm. Fig. 2(b) shows that the $LP_{01}$ modes of the four types of fibers are concentrated in the fiber core. As a representative example, we prefer to explore the NEARF performance at $\lambda = 1.06$ μm. Among the four different structures, the loss of NEARF1 at $\lambda = 1.06$ μm is the lowest at $6.45\times10^{-6}$ dB/km. The trend of the loss curve of NEARF1 and NEARF3 is the same, and the loss remains unchanged after a rapid decrease with the increase in wavelength. The minimum losses at $\lambda = 1.01$ μm are $4.66\times10^{-6}$ and $2.44\times10^{-}$

[5] dB/km. The loss curves of NEARF2 and NEARF4 fiber almost overlap. The loss decreases rapidly to the lowest point, and then increases sharply with the increase in wavelength; the minimum loss is $2.8 \times 10^{-5}$ dB/km at $\lambda = 0.96$ μm. The loss changes of the four structures satisfy the antiresonance law [29]. The lowest loss of NEARF1 is one order of magnitude lower than that of the three other fibers with different structures. The four types of fibers with different nested tube structures are optimized at $\lambda = 1.06$ μm, but the lowest loss is not obtained, and the lowest loss point occurs at a shorter wavelength. This finding shows that for the nested design of the fiber structure, to obtain the minimum loss at the predicted wavelength, considering a larger silica strut thickness may be possible to move this minimum toward the selected target wavelength.

*3.2 Bending Loss*

Bending loss is another important characteristic of fiber. Therefore, simulating the bending loss of different optical fiber structures is necessary. The results are shown in Fig. 3. We used the conformal transformation method [30] to calculate the bending loss. The bending loss is solved by formula (1), as follows:

$$n'(x, y) = n(x, y) e^{(x/R_b)} \tag{1}$$

where $n'(x, y)$ is the effective refractive index of the fiber after bending, $n(x, y)$ is the refractive index profile of the straight fiber, $R_b$ is the bending radius, and $x$ is the transverse distance from the center of the fiber.

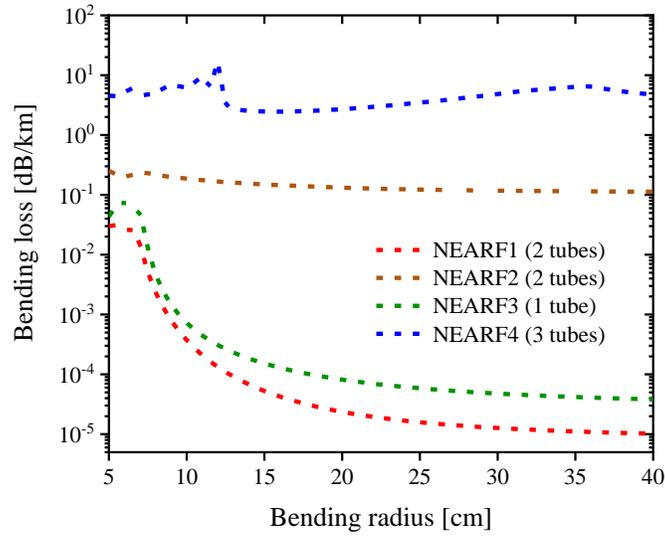

(a)

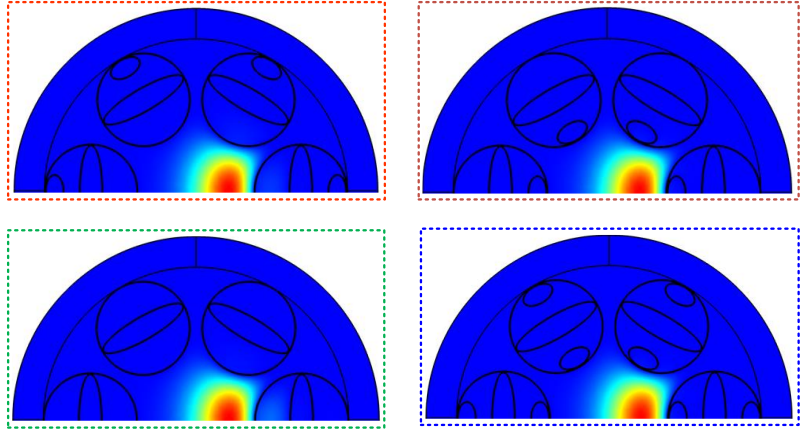

(b)

Fig. 3. Bending loss as function of wavelength for different fiber structures. (a) Bending loss curves of NEARF and (b) mode field with four cladding structures at a bending radius of 5 cm. The line color of curve relates with the color of frame of geometry.

The bending loss was calculated at $\lambda = 1.06$ μm, and the bending direction was selected as the *x*-axis giving the bend-loss curve shown in Fig. 3(a). The bending loss of the four types of fibers with nested elements are almost constant for reasonable bend radii. The reason for the bend loss peak of NEARF4 fiber may be that the core mode is coupled with the mode supported by other parts of the cladding. Within the bending radius of 5–40 cm, the bending loss of NEARF1 is lower than that of the three other fibers. The loss of NEARF1 is as low as $2.99 \times 10^{-2}$ dB/km at a bending radius of 5 cm. Fig. 3(a) also confirms that the NEARF1 has a bend loss below $1.01 \times 10^{-5}$ dB/km at 40 cm bending radius, which is approximately four and five orders of magnitude lower than that of the NEARF2 and NEARF4 fibers, respectively. The trend of the bending loss curve of NEARF1 and NEARF3 is the same, and the loss remains basically unchanged after a rapid decrease with the increase in bending radius. This finding is caused by the offset of the bending mode field that decreases, and the influence on the transmission loss of the fiber is also reduced. Fig. 3(b) shows the mode field of the bent four types of fibers with nested elements at a bending radius of 5 cm. The spot position in the figure indicates that the mode field of the curved NEARF fibers shift in the bending direction, but the mode field is completely confined in the fiber core.

### *3.3 Single-Mode Performance*

The higher-order-mode (HOM) extinction ratio (HOMER) can be used to judge the single-mode performance of HC-NCFs. The so-called HOMER is the ratio between the transmission loss of the HOM having the lowest transmission loss and the transmission loss of the FM under the same structure parameters. The HOMER simulation results of four types of fibers with nested elements are shown in Fig. 4.

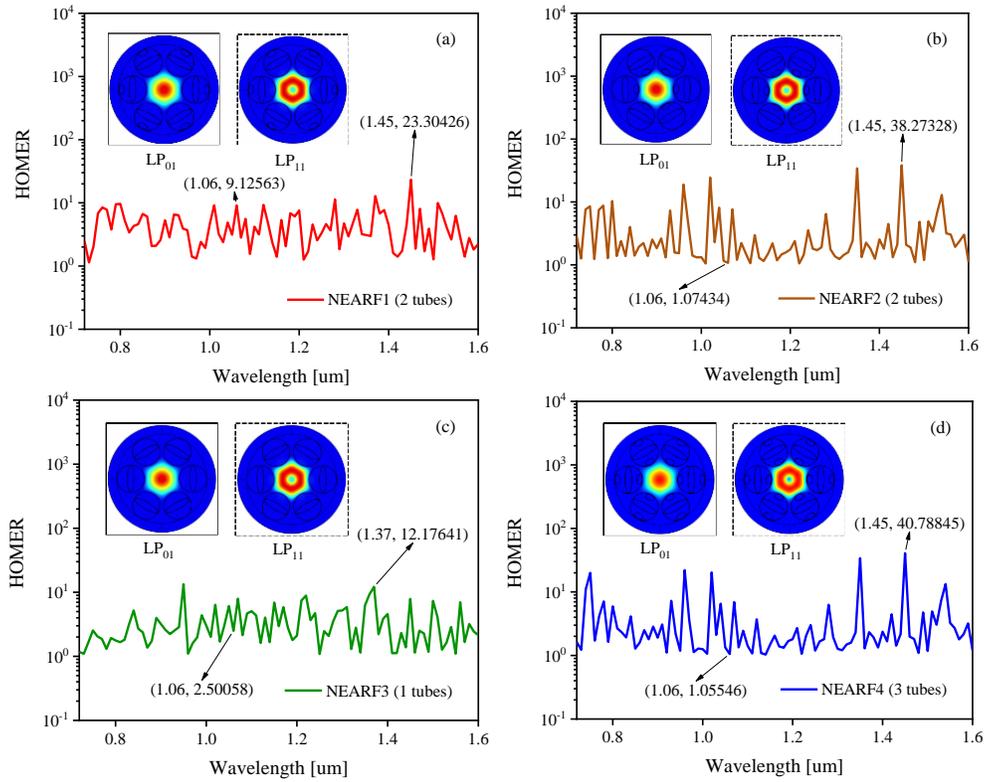

Fig. 4. HOMER as function of wavelength for all different fiber structures. The inset shows the mode fields of the $LP_{01}$ mode and the $LP_{11}$ mode.

The HOMER curves in Fig. 4 show that NEARF1 has the best single-mode performance at 1.06 μm, and its HOMER value (9.12) is higher than that of the other three fibers. In the range 0.72–1.6 μm, four types of fibers with nested elements have been optimized to provide the highest HOMER value. Among them, the highest HOMER of NEARF4 fiber is 40.78 at λ = 1.45 μm. NEARF1, NEARF2, and NEARF3 have the largest HOMER at λ = 1.45, 1.45, and 1.37 μm, respectively, and the corresponding HOMER values are 23.30, 38.27, and 12.17. The inset of Fig. 4 shows that the mode fields of the $LP_{01}$ mode and the $LP_{11}$ mode are concentrated in the HC-NCF fiber core, indicating that the $LP_{01}$ mode and the $LP_{11}$ mode are hardly coupled with the HOM supported by the cladding tube.

None of the four types of fibers have very good single-mode performance, but researchers have developed a strong interest in HOM, such as $LP_{11}$ mode. NEARF may support ultralow transmission loss of the $LP_{11}$ mode and become a dual-mode fiber with excellent performance. To this end, we calculate HOMER between another lowest loss HOM ($LP_{02}$) and the $LP_{11}$ mode of NEARF1, as shown in Fig. 5.

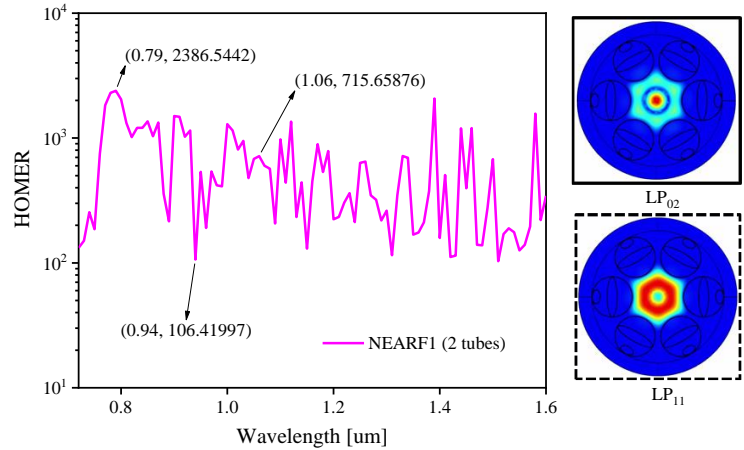

Fig. 5. HOMER between LP$_{02}$ mode and LP$_{11}$ mode (left). The mode field diagram (right) for two HOMs (LP$_{02}$ and LP$_{11}$).

Fig. 5 shows that the HOMER between the two HOMs LP$_{02}$ and LP$_{11}$ is relatively high in the range of 0.72–1.6 μm, and the HOMER value can be over 106.41. The highest HOMER of approximately 2386.54 is obtained at 0.79 μm, and the HOMER of NEARF1 can be over 715.65 at $\lambda$ = 1.06 μm. In addition, the mode field shows that the LP$_{11}$ mode only exists inside the fiber core, indicating that the introduced nested tubes ensure the low loss of the core mode and greatly increase the HOMER. Therefore, we believe that NEARF1 has a good inhibitory effect on the other HOMs.

## 4. Conclusion

In conclusion, in the same simulation environment, we use the finite element method to simulate and calculate the transmission loss, bending loss, and single-mode performance of the four types of fibers with different nested tube structures. The results show that the transmission loss of NEARF1 has better transmission performance than that of the three other fibers with different nested structures in the range of $\lambda$ = 0.72–1.6 μm. The transmission loss as low as $6.45\times10^{-6}$ dB/km is predicted at $\lambda$ = 1.06 μm. To the best of our knowledge, the record low level of transmission loss of HC-NCFs with nested tube structures was created. In addition, NEARF1 has been confirmed to have better bending resistance. Within a bending radius of 5–40 cm, NEARF1 has a bend loss below $2.99\times10^{-2}$ dB/km at 5 cm bending radius; the bend loss is below $1.01\times10^{-5}$ dB/km at 40 cm bending radius. Subsequently, we have studied the single-mode properties of four types of fibers with different nested tube structures. The NEARF1 has the best single-mode performance compared with the three other fibers at the designed $\lambda$ = 1.06 μm, and the HOMER between LP$_{11}$ mode and LP$_{01}$ mode is higher than 9.12. Its single-mode performance should be optimized. Finally, we calculate HOMER between another lowest loss HOM (LP$_{02}$) and the LP$_{11}$ mode, and the HOMER of NEARF1 can be over 106.41 in the range 0.72–1.6 μm. Therefore, NEARF1 may support ultralow transmission loss of the LP$_{11}$ mode and become a dual-mode fiber with excellent performance. We believe that the proposed NEARF1 design offers significant advantages and improvements over the designs reported in the current literature, promising excellent performance for transmission applications in the near-IR spectral regions.

**Funding.** The research was funded by the National Natural Science Foundation of China (61871031, 61875012, 61905014); Beijing Municipal Natural Science Foundation (4222017).

**Disclosures.** The authors declare no conflicts of interest.